\documentclass[aps,twocolumn,showpacs,preprintnumbers,amsmath,amssymb]{revtex4}
\usepackage{epsfig} 
\usepackage{amsmath,amssymb,amsfonts,bbm} 
\usepackage{float}%,a4wide} 
\usepackage{graphicx}% Include figure files
\usepackage{dcolumn}% Align table columns on decimal point
\usepackage{bm}% bold math
\newcommand{\bi}{\bigskip}
\newcommand{\no}{\noindent}
\newcommand{\be}{\begin{eqnarray}}
\newcommand{\ee}{\end{eqnarray}}
\newcommand{\hk}{\hspace{0.1cm}}

\newcommand{\rk}{\right)}
\newcommand{\lk}{\left(}

\newcommand{\sli}{\sum\limits}

\begin{document}

\title{Center Vortices and Ghosts}
\bigskip
\author{H. Reinhardt}
 \affiliation{Institut f\"ur Theoretische Physik\\
Auf der Morgenstelle 14\\
D-72076 T\"ubingen\\
Germany}%Lines break automatically or can be forced with \\\bigskip

\date{\today}
\bi

\no

%\maketitle
\begin{abstract}
Assuming that center vortices are the confining gauge field configurations, we
argue that in gauges that are sensitive to the confining center vortex
degrees of freedom, and where the latter lie on the Gribov horizon, the
corresponding ghost form factor is infrared divergent. Furthermore, 
this infrared divergence
disappears when center vortices are removed from the Yang-Mills ensemble. On
the other hand, for gauge conditions which are insensitive to center vortex
degrees of freedom, the ghost form factor is infrared finite and does not change
(qualitatively) when center vortices are removed. 
Evidence for our observation is provided 
from lattice calculations.
\end{abstract}
%\date{\today}

\bi

\no
%Pfad: paper/unpublished/paper-centervortices/centervortices.tex
\maketitle

\bi

\no
\section{Center dependent gauges}
\bi

\no
In Landau gauge the Kugo-Ojima confinement criterium requires an infrared
divergent ghost form factor. Indeed, the ghost form factor in Landau gauge has
been calculated on the lattice and is found to be infrared divergent 
\cite{Gattnar:2004bf},
\cite{Ilgenfritz:2006gp}.
Furthermore, the lattice calculations also show, when the confining center
vortex field configurations detected by the method of center projection
\cite{DelDebbio:1996mh} are
removed from the Yang-Mills ensemble (by the method of ref.
 \cite{deForcrand:1999ms}), the
temporal string tension disappears and, at the same time, the ghost form factor
of Landau gauge looses its infrared divergent behaviour 
\cite{Gattnar:2004bf}, see fig. 1,
which is a necessary and sufficient condition for the Kugo-Ojima confinement
criterium \cite{Kugo:1979gm} to be realized.
\begin{figure}[t]
\begin{center}
\epsfig{file=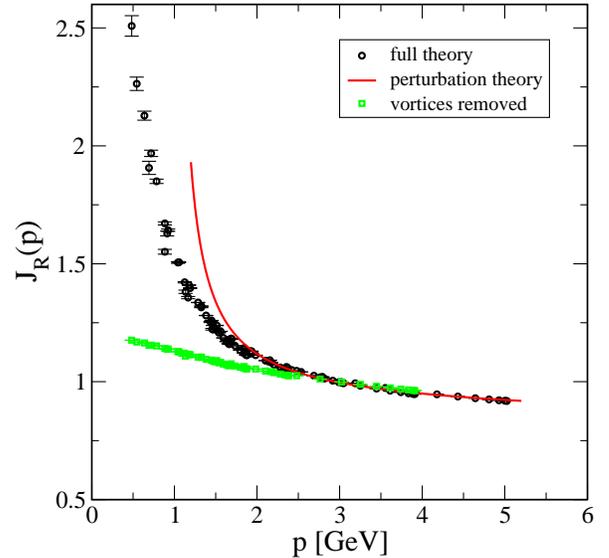,height=7.4cm,clip}
\end{center}
\caption{{\sl (taken from ref.1) 
The renormalized ghost form factor as function of the momentum
transfer $p$ for full SU(2) gauge theory and for the center vortex removed
ensemble,
respectively.}
\label{}}
\end{figure}
Thus, there seems to be an intrinsic relation between the Kugo-Ojima
confinement criterium in Landau gauge and Wilson's confinement criterium, i.e. an
area law for the temporal Wilson loop. However, recent lattice calculations 
\cite{R6}
 show that the ghost form factor is infrared divergent even above the
critical temperature in the deconfined region. Thus, an infrared divergent ghost
form factor is a necessary but not yet sufficient condition for confinenent (in
the sense of Wilson) in
Landau gauge. 
\bi

\no
According to the confinement mechanism proposed by Gribov \cite{Gribov:1977wm} 
and further
elaborated by Zwanziger \cite{Zwanziger:1998ez} 
confinement arises due to the infrared dominance of the
field configuration near the Gribov horizon, which are expected to give rise to
an infrared diverging ghost form factor. This confinement scenario is expected
to be realized in both Landau and Coulomb gauge. 
In Coulomb gauge, in particular, an infrared divergent
ghost form factor is required for a confining (i.e. linearly arising) static
Coulomb potential, which is a necessary but not sufficient condition 
\cite{Zwanziger:2002sh} 
for confinement in the sense of Wilson's criterium.
Gribov's confinement scenario and the center vortex picture of confinement are
compatible in the sense that center vortices lie on the Gribov horizon in both
Landau and Coulomb gauge \cite{Greensite:2004ke} 
and the corresponding ghost form factors loose their
infrared diverging behaviour, when center vortices are removed. This has been
explicitely demonstrated in Landau gauge \cite{Gattnar:2004bf} (see fig. 1) 
and, on the basis of the results obtained in ref. \cite{Greensite:2004ke}, 
can be expected to be true also in
Coulomb gauge.
\bi

\no
Consider Landau and Coulomb gauge on the lattice defined by 
\be
\label{1}
\sli_x \sli^d_{\mu = 1} tr U_\mu (x) \to max \hk ,
\ee
where $d = D$ for Landau gauge and $d = D - 1$ for Coulomb gauge ($D-$number of
space time dimensions). Eliminating the center vortices by the method of ref.
\cite{deForcrand:1999ms} 
implies to multiply the links $U_\mu (x)$ in the so-called
maximal center gauge (see below)
by $Z_\mu = sign (tr U_\mu)$. This procedure changes the gauge condition
(\ref{1}) to
\begin{multline}
\label{2}
\sli_x \sli^d_{\mu = 1}  tr U_\mu (x) sign (tr U_\mu (x))
 \\
=\sli_x \sli^d_{\mu = 1} | tr U_\mu
(x)| \to max \hk .
\end{multline}
The original Landau and Coulomb 
gauge conditions (\ref{1}) and consequently also the
corresponding ghost Green functions obviously feel the confining center vortex
degrees of freedom. Eliminating the confining center vortices turns the 
gauge condition (\ref{1}) in the gauge condition (\ref{2}), which does no longer
depend on the center vortex degrees of freedom, and, hence, the corresponding
ghost form factor is insensitive to the confining center vortices. The modified
gauge condition (\ref{2}) is for $d = D$, 
in fact, equivalent to the so-called maximum
center gauge condition.
\bi

\no
\section{Center independent gauges}
\bi

\no
The maximal center gauge is defined by the condition \cite{DelDebbio:1996mh}
\be
\label{3}
\sli_x \sli^D_{\mu = 1} (tr U_\mu (x))^2 \to max \hk .
\ee
This condition fixes the gauge group $SU (2)$ only up
to the coset $SU (2) / Z (2) \simeq SO (3)$ and is thus insensitive to the
center vortex degrees of freedom (the replacement $U_\mu \to Z_\mu U_\mu$
obviously does not change the gauge condition). In fact, using the $SU (2)$
trace identity $2 (tr U)^2 = tr \hat{U} + 1$, where $\hat{U}^{a b} =
\frac{1}{2} tr \lk \tau_a U \tau_b U^\dagger \rk$ is the
adjoint representation of $U$, eq. (\ref{3}) becomes
\be
\label{4}
\sli_x \sli^D_{\mu = 1} tr \hat{U}_\mu (x) \to max  \hk ,
\ee
which manifestly depends only on the coset variables $\hat{U}_\mu \in SU (2) / Z
(2) \simeq SO (3)$. The maximal center gauge condition (\ref{3}) is, in fact,
equivalent to the condition (\ref{2}) following from the Landau gauge by
eliminating the center degrees of freedom as is also seen from eq. (\ref{4}).
\bi

\no
In the continuum theory it is explicitly seen that the maximal 
center gauge brings a given
gauge configuration as close as possible to the ``nearest'' center vortex
configuration, and reduces to the Landau
gauge in the absence of center vortices in the gauge configuration considered
\cite{Engelhardt:1999xw}.
Furthermore, for gauge configurations which are pure center vortices, the
maximal center gauge condition is identically fulfilled and thus does not fix
the gauge of center vortex configurations at all
(see eq. (76) of ref. \cite{Engelhardt:1999xw}).
\bi

\no
Since the maximal center gauge does not feel the confining center vortex degrees
of freedom, one expects that its ghost form factor is infrared finite (and
consequently does not change by a removal of center vortices). This, indeed, is
found in lattice calculations, ref. \cite{Langfeld:2005kp}, see figure 2.
\begin{figure}[t]
\begin{center}
\epsfig{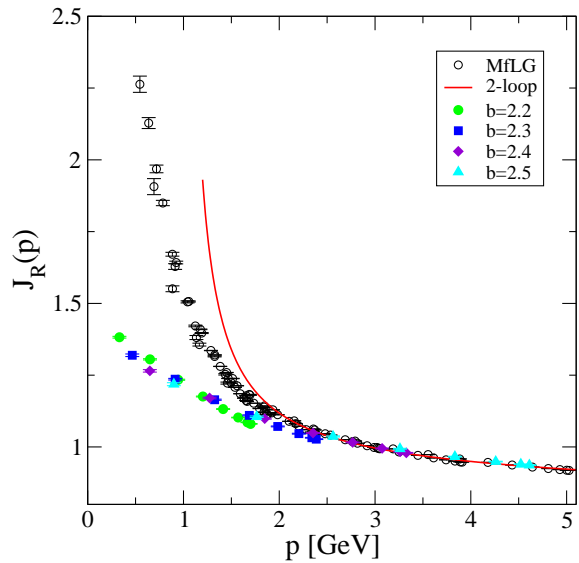}
%\vspace{-5mm}
\caption{{\sl (taken from ref.11)
 The ghost form factor (full symbols) in maximal center gauge as function of
momentum. For sake of comparison the ghost form factor in Landau gauge (\ref{1})
and in two loop perturbation theory are also shown.
}}
\label{ghost}
\end{center}
\end{figure}
\bi

\no
Abelian dominance and evidence for the dual Meissner effect was most
significantly observed on the lattice by using the method of Abelian projection
in the maximal Abelian
 gauge. The maximal Abelian gauge is defined on the lattice
by
\be
\label{5}
\sli_{x, \mu} \frac{1}{2} tr \lk \tau_3 U_\mu (x) \tau_3 U^\dagger_\mu (x) \rk =
\sli_{x, \mu} \lk \hat{U}_\mu (x) \rk^{33} \to max \hk 
\ee
and fixes only the coset $SU (2) / U (1)$. It was expected, that the ghost form 
factor 
in the maximal Abelian gauge behaves
similar to the one in Landau gauge \cite{R12}, 
since the maximal Abelian gauge implies
Landau gauge for the Abelian projected part of the gauge field. However, like
 the maximal center gauge, also
this gauge condition does not feel the confining 
center degrees of freedom. Consequently,
its ghost form factor is expected to be infrared finite,too. Indeed, this is
observed in recent lattice calculations \cite{R12}. Since the maximal
Abelian gauge condition depends only on the coset degrees of freedom $SU (2) / U
(1)$ and thus does not feel the center $Z (2) \subset U (1)$, 
we also expect that the ghost form factor does not change by removing the
center vortices.
\bi

\no
\section{Conclusions}
\bi

\no
The above considered examples shows: If a gauge condition is sensitive to the
confining center vortex degrees of freedom and the latter lie on the Gribov 
horizon (like in Landau gauge and Coulomb gauge),
the corresponding ghost form factor is infrared divergent. This infrared
divergence disappears, however, when the center vortices are removed from the
Yang-Mills ensemble. 
\bi

\no
On the other hand, if a gauge conditon is insensitive to the center vortex
degrees of freedom (like 
maximal center gauge and maximal Abelian gauge) its ghost
form factor is infrared finite and does not qualitatively 
change, when the center vortices are
removed.
 \bi

\no
Acknowledgement:
\bi

\no
Discussions with G. Burgio, A. Cucchieri, T. Mendes, M. Quandt and P. Watson
 are gratefully acknowledged.
The author is grateful to A. Cucchieri and T. Mendes for providing their
lattice results prior to publication. This work was supported by DFG Re 856/4-3.
\bi

\no

\end{document}